\documentclass[lettersize,journal]{IEEEtran}

\usepackage[T1]{fontenc}
\usepackage[utf8]{inputenc} 
\usepackage{graphicx}
\usepackage[]{hyperref}
\usepackage[version=3]{mhchem}
\usepackage{amssymb}
\usepackage{longtable}
\usepackage{multirow}
\usepackage{amsmath}
\usepackage{eurosym}	 
\usepackage{mathtools}
\usepackage{booktabs}

\usepackage{upgreek}

\usepackage[T1]{fontenc}
\usepackage[version=3]{mhchem}
\usepackage{amssymb}
\usepackage{longtable}
\usepackage{multirow}
\usepackage{amsmath}
\usepackage{mathtools}
\usepackage{epstopdf}
\usepackage{caption}
\usepackage{color}
\usepackage{xcolor}
\definecolor{myred}{rgb}{0.7,0.15,0.15}
\definecolor{mygreen}{rgb}{0.13,0.55,0.13}
\definecolor{myblue}{rgb}{0.25,0.41,0.88}

\usepackage{graphicx}
\usepackage[labelformat=simple]{subcaption} 

\usepackage{placeins}
\usepackage[font={small}]{caption}

\usepackage{mathrsfs}
\usepackage{comment}

\usepackage{enumitem}
\usepackage{url}
\usepackage{times}

\usepackage{pgfplots}
\usepackage{tikz}
\usetikzlibrary{patterns}
\usetikzlibrary{external}
\usetikzlibrary{spy}

\usepackage{pgfkeys}
\usetikzlibrary{intersections}
\usepackage{listings}

\makeatletter
\newcommand\Label[1]{&\refstepcounter{equation}\text{(\theequation)}\ltx@label{#1}&}
\makeatother

\newlength\Colsep
\renewcommand{\vec}[1]{\boldsymbol{#1}}

\newcommand{\paren}[1]{\left(#1\right)}

\renewcommand{\b}{\vec b}

\newcommand{\n}{\vec n}
\newcommand{\h}{\vec h}

\newcommand{\mur}{\mu_\text{r}}
\newcommand{\e}{\vec e}

\renewcommand{\j}{\vec j}

\newcommand{\ec}{e_{\text{c}}}
\newcommand{\jc}{j_{\text{c}}}
\newcommand{\bs}{\vec b_{\text{s}}}

\newcommand{\hpf}{$h$-$\phi$-formulation\ }

\newcommand{\hbf}{$h$-$\phi$-$b$-formulation\ }
\newcommand{\ajf}{$a$-$j$-formulation\ }

\newcommand{\hpfOnly}{$h$-$\phi$-formulation}

\newcommand{\hbfOnly}{$h$-$\phi$-$b$-formulation}

\newcommand{\tafOnly}{$t$-$a$-formulation}

\newcommand{\hpfc}{$h$-$\phi$\ }

\newcommand{\hbfc}{$h$-$\phi$-$b$\ }
\newcommand{\ajfc}{$a$-$j$\ }

\newcommand{\hpfcOnly}{$h$-$\phi$}

\newcommand{\hbfcOnly}{$h$-$\phi$-$b$}
\newcommand{\ajfcOnly}{$a$-$j$}

\usepackage{eso-pic}

\usepackage{color}
\usepackage{xcolor}
\definecolor{myred}{rgb}{0.7,0.15,0.15}
\definecolor{mymaincolor}{rgb}{0.24, 0.36, 0.64}
\definecolor{mysecondcolor}{rgb}{0.21, 0.64, 0.87}
\definecolor{myblue}{rgb}{.2,0.45,0.5} 
\definecolor{myorange}{rgb}{0.78,0.6,0.3}
\definecolor{mygreen}{rgb}{.2,0.38,0.16}
\definecolor{myalert}{rgb}{0.97,0.09,0.21}

\definecolor{myformulation}{rgb}{0.33, 0.29, 0.31}
\definecolor{myformulation_back}{rgb}{1, 0.97, 0.91}

\definecolor{hf}{rgb}{0.93, 0.57, 0.13} 
\definecolor{hf_2}{rgb}{1.0, 0.89, 0.77} 
\definecolor{hf_3}{rgb}{1.0, 0.22, 0.0} 
\definecolor{hf_4}{rgb}{1.0, 0.4, 0.1} 

\definecolor{burlywood}{rgb}{0.87, 0.72, 0.53}
\definecolor{burntorange}{rgb}{0.8, 0.33, 0.0}
\definecolor{burntsienna}{rgb}{0.91, 0.45, 0.32}

\definecolor{af}{rgb}{0.4, 0.53, 0.34}
\definecolor{af_2}{rgb}{0.74, 0.77, 0.47}
\definecolor{af_3}{rgb}{0.12, 0.3, 0.17}
\definecolor{af_4}{rgb}{0.03, 0.34, 0.25}

\definecolor{haf}{rgb}{0.6, 0.51, 0.48}
\definecolor{haf_2}{rgb}{1, 0.97, 0.91}

\definecolor{taf}{rgb}{0, 0.55, 0.5}
\definecolor{ajf}{rgb}{0.29, 0.59, 0.82}

\definecolor{hbf}{rgb}{0.87, 0.36, 0.51}

\begin{document}

\title{Finite Element Models for Magnetic Shields Made of Stacked Superconducting Tape Annuli}

\author{Julien~Dular, Sébastien~Brialmont, Philippe~Vanderbemden,
        Christophe~Geuzaine,~and~Benoît~Vanderheyden
\thanks{J. Dular is with CERN, TE-MPE-PE, Geneva, Switzerland. The other authors are with the Department of Electrical Engineering and Computer Science, Institut Montefiore B28 at the University of Liège, B-4000 Liège, Belgium.}%
}

\maketitle

\begin{abstract}
Stacks of high-temperature superconducting tape annuli can be used as magnetic shields operating efficiently for both axial and transverse fields. However, due to their layered geometry and hybrid electrical and magnetic properties, implementing models of such structures is not straightforward. In this work, we propose two different modelling approaches with the finite element method: layered and homogenized. We compare their accuracy and numerical efficiency for three different formulations (\hpfcOnly, \hbfcOnly, and \ajfcOnly), in both axial (2D-axisymmetric) and transverse (3D) configurations. We show that both approaches lead to comparable performance in the axial case, but that the homogenized model is considerably harder to use in the transverse case.
\end{abstract}


\IEEEpeerreviewmaketitle


\AddToShipoutPicture*{
    \footnotesize\sffamily\raisebox{0.8cm}{\hspace{1.5cm}\fbox{
        \parbox{\textwidth}{
            This work has been submitted to the IEEE for possible publication. Copyright may be transferred without notice, after which this version may no longer be accessible.
            }
        }
    }
}

\section{Introduction}
\addcontentsline{toc}{section}{Introduction}

\IEEEPARstart{C}{ompared} to conventional magnetic shields made of ferromagnetic materials only, superconducting shields are not limited by a saturation magnetization and can therefore operate at much higher flux densities~\cite{Durrell2018,brialmont2022magnetic}. Shielding devices may involve low-temperature supeconductors \cite{takahata1989magnetic,sasaki1996magnetic} or high-temperature superconductors (HTS). For HTS, bulk materials~\cite{denis2007magnetic, yang2017fabrication,fagnard2019magnetic} or coated conductors can be used, such as eye-shaped loops~\cite{fagnard2010use,wera2015comparative} or stacks of tapes~\cite{peng2021passive}.

\begin{figure}[h!]
            \centering 
		\includegraphics[width=0.86\linewidth]{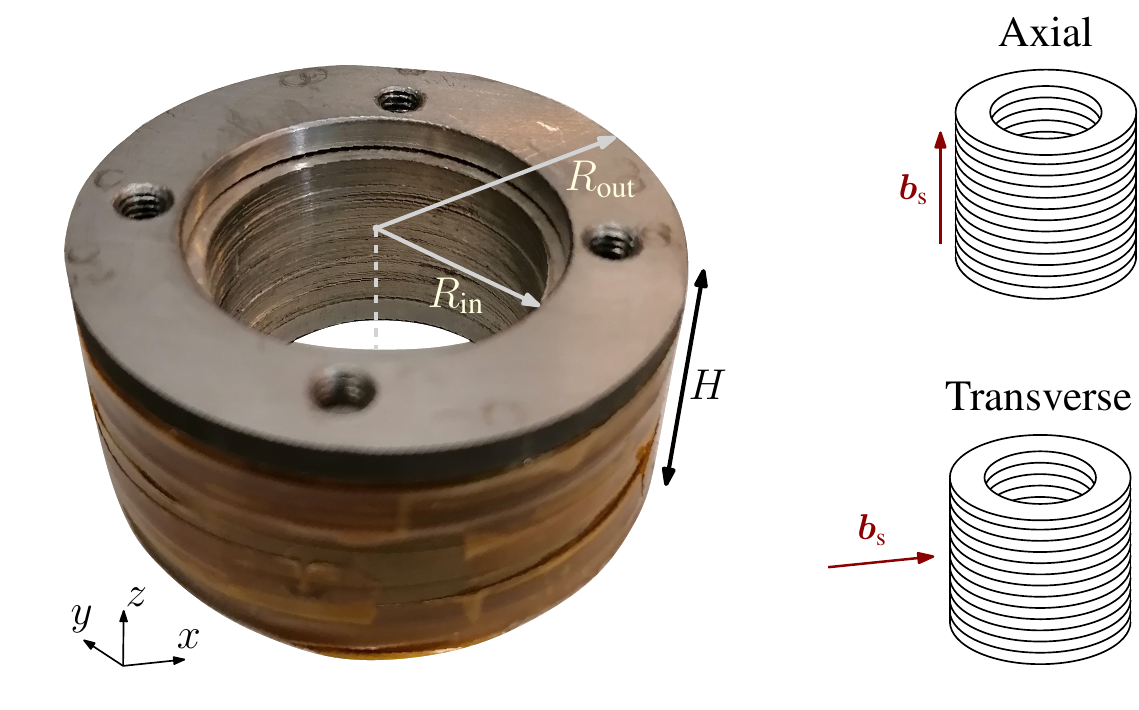}
\caption{Magnetic shield made up of $N=182$ stacked tape annuli. (Left) Sample used for experimental measurements described in~\cite{brialmont2022magnetic}. (Right) Illustration of the axial and transverse configurations.}
\label{stackedTape_geometry}
\end{figure}

We focus on the last category: stacks of HTS tapes. We consider a shield that consists of a stack of a large number of YBCO tape annuli with a large bore (26~mm), as illustrated in Fig.~\ref{stackedTape_geometry}. More specifically, we study the finite element modelling of its magnetic response at 77 K.

In previous work~\cite{brialmont2022magnetic}, we demonstrated experimentally the excellent shielding ability of such stacks and modelled the corresponding magnetic shielding properties using an \hpf with an homogenized model. The goal of the present work is to make a thorough comparison of the numerical performance of alternative modelling approaches.

Conducting this numerical study is relevant for at least two reasons. First, the layered structure of the stack of tapes makes the geometry of the problem very intricate. It is however not obvious whether such a detailed description of the geometry is necessary for extracting reliable predictions on the shielding properties of the system. For this reason, we propose and compare the predictions of two distinct models: a layered (detailed) model, and an homogenized model~\cite{gyselinck2004time}.

Second, as a significant volume fraction of each tape consists in a ferromagnetic (FM) substrate, the whole structure is a HTS-FM hybrid, having field-dependent resistivity and permeability. Choosing an efficient numerical method for these hybrid structures is not trivial as different approaches may result in completely different numerical behaviors~\cite{dular3Dmagnet}.

The paper is structured as follows. After defining the shielding problem in Section~\ref{sec_definition}, we describe the layered and homogenized models in Sections~\ref{sec_layeredmodel} and \ref{sec_homogenizedmodel}, respectively. We then assess the accuracy and efficiency of both models when solved with three different finite element formulations: the \hpfcOnly, \hbfcOnly, and \ajfcOnly-formulations~\cite{Stenvall2010a, dular2023standard}: in Section~\ref{sec_resultAxial}, we consider the shielding with respect to an axial field, which can be modelled as a 2D-axisymmetric problem, and in Section~\ref{sec_resultTransverse}, we analyze a transverse field configuration, which requires a 3D model. The objective of the analysis is to provide recommendations for obtaining accurate and efficient numerical resolutions. 

Models are solved with GetDP~\cite{getdp}, geometry and mesh generation is performed with Gmsh~\cite{gmsh} and codes are based on the Life-HTS toolkit\footnote{Available online: \url{www.life-hts.uliege.be}.}. All are open-source projects.

\section{Problem Definition}\label{sec_definition}

We consider a magnetic shield made up of a stack of $N = 182$ YBCO tape annuli, extracted from a $46$ mm-wide coated conductor~\cite{brialmont2022magnetic}. The superconducting tape is based on a rolling assisted biaxially textured substrate (RABiTS)~\cite{goyal2004rabits}, made of Ni-5at.\%W. The volume fraction of the ferromagnetic (FM) substrate in the tape, and hence also in the whole stack, is $f=0.92$. Annuli have an inner radius $R_\text{in} = 13$~mm and an outer radius $R_\text{out} = 22.5$~mm. The sample height is $H=14.9$~mm. The fabrication process is described in \cite{patel2016magnetic}.

\subsection{Shielding Factor}

We study the magnetic response of the shield to a uniform applied field $\bs$ in two configurations: axial and transverse, as represented in Fig.~\ref{stackedTape_geometry}. In the axial configuration, the field $\bs$ is parallel to the symmetry axis of the stack of tapes, i.e., in the $z$-direction. In the transverse configuration, the field is perpendicular to the cylinder axis, e.g., in the $x$-direction. To quantify the shielding properties of the stack of tapes, we define the shielding factor SF as follows:
\begin{equation}\label{eq_SF_definition}
\text{SF} = \frac{\|\bs\|}{\|\b_\text{in}\|},
\end{equation}
where $\b_\text{in}$ is the magnetic flux density measured (or computed) at the center of the stack. The shielding factor SF will be our main indicator during the numerical analysis. It is a local indicator, thus it is very sensitive to the solution quality.

Shielding factors in different situations were measured experimentally in our previous work~\cite{brialmont2022magnetic}. Here, we consider two cases, both at 77~K: (i) axial configuration, with a field ramped from $0$ to $670$ mT at a rate of $5$ mT/s, and (ii) transverse configuration, with a field ramped from $0$ to $60$~mT at a rate of $0.75$ mT/s. 

\subsection{Material Parameters}

\begin{figure}[h!]
        \centering
	\begin{subfigure}[b]{0.49\linewidth}  
     \centering 
     \tikzsetnextfilename{permeabilityMeasurement_bh_curve}
     \hspace{0.3cm}
\begin{tikzpicture}[trim axis left, trim axis right][font=\small]
 	\begin{axis}[
    width=1\linewidth,
    height=4.2cm,
    grid = both,
    grid style = dotted,
    xmin=0, 
    xmax=2000,
    ymin=0, 
    ymax=0.220,
    ytick={0, 0.05, 0.1, 0.15, 0.2},
    yticklabels={$0$, $0.05$, $0.1$, $0.15$, $0.2$},
    xtick={0, 1000, 2000},
    xticklabels={$0$, $1000$, $2000$},
	xlabel={$\|\h\|$ (A/m) },
    ylabel={$\|\b\|$ (T)},
    ylabel style={yshift=-1.1em},
    xlabel style={yshift=0.4em},
    legend style={at={(0.95, 0.2)}, anchor= east, draw=none}
    ]
    \addplot[myblue, thick] 
    table[x=h,y=b]{data_stackedTapes/stacked_permeability_77K.txt};
    \end{axis}
	\end{tikzpicture}%
	\vspace{-0.2cm}
    \end{subfigure}
        \hspace{-0.2cm}
        \begin{subfigure}[b]{0.49\linewidth}
            \centering
            \tikzsetnextfilename{permeabilityMeasurement_mur_curve}
\begin{tikzpicture}[trim axis left, trim axis right][font=\small]
 	\begin{axis}[
 	tick scale binop=\times,
    width=1.1\linewidth,
    height=4.2cm,
    grid = both,
    grid style = dotted,
    xmin=0, 
    xmax=0.350,
    ymin=0, 
    ymax=400,
	ylabel={$\mu_\text{r}$ (-) },
    xlabel={$\|\b\|$ (T)},
    ylabel style={yshift=-1.2em},
    xlabel style={yshift=0.4em},
    legend style={at={(0.99, 0.85)}, anchor= east, draw=none}
    ]
        \addplot[myblue, thick] 
    table[x=b,y=mur]{data_stackedTapes/stacked_permeability_77K.txt};
    \end{axis}
	\end{tikzpicture}%
	\vspace{-0.2cm}
        \end{subfigure}
        \caption{First magnetization curve (left) and relative permeability (right) of the FM substrate at 77~K. Below 0.2~T, interpolation of experimental data~\cite{brialmont2022measurement}; above 0.2~T, smooth extrapolation of them.}
        \label{permeabilityMeasurement}
\end{figure}
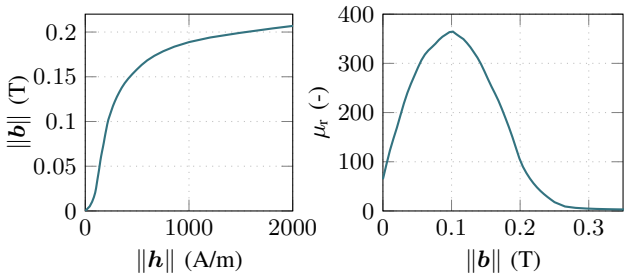

The first magnetization curve of the FM substrate was measured at 77~K~\cite{brialmont2022measurement}. We use the smooth fitting curves represented in Fig.~\ref{permeabilityMeasurement} to describe the magnetic constitutive law of the FM substrate, which occupies a volume fraction $f=0.92$ of the stack of tapes. For the complementary volume fraction, $1-f$, containing the HTS layer as well as buffer layers, we assume that $\mu = \mu_0$.

Only the superconducting layer is modelled as a conducting material. We neglect the conductivity of the FM substrate and other layers, and we assume that the HTS fills a volume fraction $1-f$ of the whole stack. We describe the electrical resistivity of the HTS layer by a power law with $n=20$ and a field-dependent critical current density $\jc(\b)$ that follows Kim's law~\cite{zhang2018study}:
\begin{align}
\rho(\j;\b) = \frac{\ec}{\jc(\b)} \paren{\frac{\|\j\|}{\jc(\b)}}^{n-1},\quad \jc(\b) = \frac{j_\text{c0}}{1+ \|\b\|/b_0},
\end{align}
with $j_\text{c0} = 7\times 10^9$ A/m$^2$ and $b_0 = 0.1$ T. For simplicity, we assume that $\jc(\b)$ only depends on the norm $\|\b\|$ of the magnetic flux density. In reality, tapes are more sensitive to fields perpendicular to their surface than to parallel fields \cite{uglietti2009angular, senatore2015field}, but we neglect this anisotropic field-dependence here.



\section{Layered Model}\label{sec_layeredmodel}

For the first model, we describe the magnetic shield with a limited number of tapes $N_{\text{s}}<N$, with fictitious thickness $w_{\text{s}} = H/N_{\text{s}}$, in order to keep the total height $H$ of the stack unchanged. As a first step in the numerical analysis, we will assess the influence of $N_{\text{s}}$ on the obtained shielding factors to verify the validity of this approach.

We assume that each fictitious tape consists of the superposition of a FM layer of thickness $fw_{\text{s}}$ and an HTS layer of thickness $(1-f)w_{\text{s}}$.

\begin{figure}[h!]
            \centering 
		\includegraphics[width=0.95\linewidth]{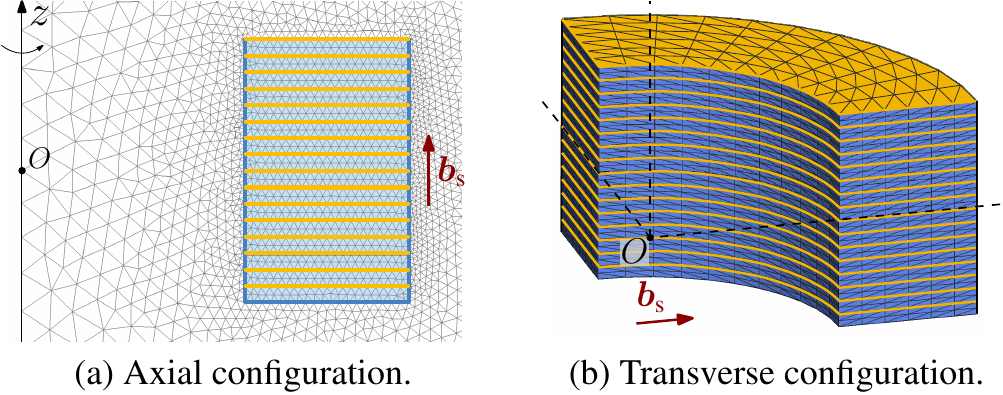}
\caption{Geometry of the layered model. Yellow regions represent the HTS layers and blue regions represent the FM layers.}
\label{geometry_presentation}
\end{figure}

In the axial configuration, the problem is axisymmetric and can be described by a 2D model, as represented in Fig.~\ref{geometry_presentation}(a). In the transverse configuration, a 3D model is necessary. By symmetry, only a quarter of the domain can be modelled, as illustrated in Fig.~\ref{geometry_presentation}(b).

The external field is imposed via an essential boundary condition on a circular or spherical external surface, placed at a distance $R=100$ mm from the center $O$. Symmetry conditions are imposed on symmetry surfaces. A global condition on either the current or the voltage is associated with each tape. In the axial configuration, screening currents are free to appear, so the voltage is fixed to zero. In the transverse configuration, by symmetry, both the current and the voltage are equal to zero in each tape.

\section{Homogenized Model}\label{sec_homogenizedmodel}

\begin{figure}[h!]
            \centering 
		\includegraphics[width=\linewidth]{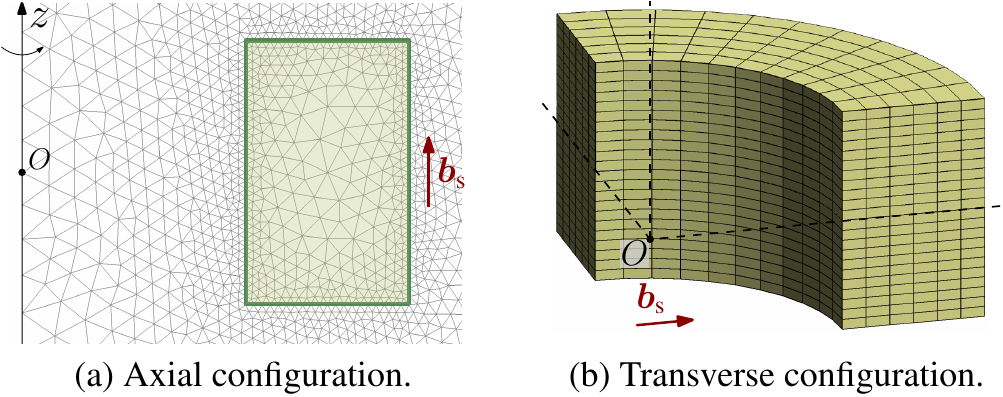}
\caption{Geometry of the homogenized model. The green region represents the hybrid HTS-FM material with anisotropic properties.}
\label{homogeneousModel}
\end{figure}

For the second method, we replace the layered structure by an homogenized material with anisotropic properties~\cite{gyselinck2004time,wang2011new,zermeno20143d}. The homogenized material has both superconducting and ferromagnetic properties, and the problem is written in terms of local averaged fields $\h$, $\b$, $\e$, and $\j$, or \textit{macroscale} fields~\cite{niyonzima2012finite}, assuming that each finite element in the homogenized model covers a sufficiently high number of tapes. The problem geometry is illustrated in Fig.~\ref{homogeneousModel} (note that a plane symmetry with respect to $z=0$ can be introduced). Boundary conditions are identical to those in the layered model, but global conditions only involve one conducting domain.

\subsection{Mesoscale Fields}

The homogenized problem is solved in terms of the macroscale fields $\h$ (or $\b$) and $\j$. However, the field-dependent permeability in the FM substrate and the power law in the HTS layer are functions of the local fields, or \textit{mesoscale} fields.

The macroscale field $\h$ is defined as the volume average $\h = f \h^\text{F} + (1-f) \h^\text{S}$, where $\h^\text{F}$ and $\h^\text{S}$ are the local fields in the FM and HTS materials, respectively. By continuity of $\h\times \n$ and $\b\cdot \n$, with $\n$ the normal of the interface, we have
\begin{align}
h_x^\text{F} = h_x^\text{S},\qquad h_y^\text{F} = h_y^\text{S},\qquad \mu(\h^\text{F}) h_z^\text{F} = \mu_0 h_z^\text{S},
\end{align}
in terms of the components of $\h^\text{F}$ and $\h^\text{S}$ in the Cartesian coordinate system represented in Fig.~\ref{stackedTape_geometry}. Consequently, the relation between the components of $\h^\text{F}$ and those of $\h$ reads
\begin{align}\label{eq_homog_h_to_hf}
\begin{pmatrix}
h_x^\text{F}\\
h_y^\text{F}\\
h_z^\text{F}
\end{pmatrix} = \begin{pmatrix}
h_x\\
h_y\\
\mu_0 h_z/\paren{f \mu_0 + (1-f)\mu(\h^\text{F})}
\end{pmatrix}.
\end{align}
Because of the nonlinearity of $\mu(\h^\text{F})$, expressing $\h^\text{F}$ in terms of $\h$ therefore requires to solve an implicit equation for $h_z^\text{F}$ at each point where the permeability value is needed. We solve this equation using a quasi-Newton method in which the Jacobian is approximated by a finite difference.

Following similar steps as for the magnetic field, we can express the relation between the Cartesian components of the macroscale magnetic flux density, $\b$, and those of the mesoscale field in the FM, $\b^\text{F}$, as follows:
\begin{align}\label{eq_homog_b_to_bf}
\begin{pmatrix}
b_x^\text{F}\\
b_y^\text{F}\\
b_z^\text{F}
\end{pmatrix} = \begin{pmatrix}
\nu_0 b_x/\paren{f \nu_0 + (1-f)\nu(\b^\text{F})}\\
\nu_0 b_y/\paren{f \nu_0 + (1-f)\nu(\b^\text{F})}\\
b_z
\end{pmatrix}.
\end{align}
This also involves an implicit equation for the $x$ and $y$-components of $\b^\text{F}$, that we also handle with a quasi-Newton method.

For the current density, because the FM substrate is assumed non-conducting, the relation between the macroscale current density $\j$ and the mesoscale current density $\j^\text{S}$ in the HTS layer is simply given by $\j^\text{S} = \j/(1-f)$. 

\subsection{Anisotropic Permeability and Reluctivity}

The homogenized material is anisotropic and its magnetic permeability (e.g. used in the \hpfOnly) takes the form of a diagonal tensor, defined as follows~\cite{gyselinck2004time}:
\begin{align}\label{eq_homog_perm_tensor}
&\tilde{\vec \mu}(\vec h^\text{F}) = \begin{pmatrix}
\bar \mu(\h^\text{F}) & 0 & 0\\
0 & \bar \mu(\h^\text{F}) & 0\\
0 & 0 & \bar{\bar{\mu}}(\h^\text{F})
\end{pmatrix}\notag\\
&\qquad \text{with}\quad  \left\{\begin{aligned}
\bar \mu(\h^\text{F}) &= f\mu(\h^\text{F})+(1-f)\mu_0,\\
\bar{\bar{\mu}}(\h^\text{F}) &= \paren{f/\mu(\h^\text{F}) + (1-f)/\mu_0}^{-1},
\end{aligned}
\right.
\end{align}
where $\h^\text{F}$ is computed from $\h$ using Eq.~\eqref{eq_homog_h_to_hf}.

Conversely, for formulations that involve the magnetic flux density as a primal unknown, e.g., the \hbfc and \ajfcOnly-formulations, the magnetic reluctivity takes the form of a diagonal tensor as well, defined as follows:
\begin{align}\label{eq_homog_reluctivity_tensor}
&\tilde{\vec \nu}(\b^\text{F}) = \begin{pmatrix}
\bar \nu(\b^\text{F}) & 0 & 0\\
0 & \bar \nu(\b^\text{F}) & 0\\
0 & 0 & \bar{\bar{\nu}}(\b^\text{F})
\end{pmatrix}\notag\\
&\qquad\text{with}\quad  \left\{\begin{aligned}
\bar \nu(\b^\text{F}) &= \paren{f/\nu(\b^\text{F}) + (1-f)/\nu_0}^{-1},\\
\bar{\bar{\nu}}(\b^\text{F}) &= f\nu(\b^\text{F})+(1-f)\nu_0,
\end{aligned}
\right.
\end{align}
where $\b^\text{F}$ is computed from $\b$ using Eq.~\eqref{eq_homog_b_to_bf}.

\subsection{Anisotropic Resistivity}

The current density can only flow in the $(x,y)$-plane. In the axial configuration, the current density is azimuthal by construction. The isotropic power law can be used as is.

In the transverse configuration, to prevent current from flowing in the $z$-direction, we introduce a relatively large resistivity $\rho_\infty$ in that direction~\cite{zermeno20143d}. In practice, $\rho_\infty = 0.01$~$\Omega$m gives satisfying results here. In the Cartesian coordinate system represented in Fig.~\ref{stackedTape_geometry}, the resulting resistivity takes the form of a diagonal tensor, defined as follows:
\begin{align}\label{eqn_stackedTapes_homogenizedResistivity}
\tilde{\vec \rho}(\j^{\text{S}};\b^{\text{S}}) =\frac{1}{1-f}\ \begin{pmatrix}
\rho(\j^\text{S};\b^{\text{S}}) & 0 & 0\\
0 & \rho(\j^\text{S};\b^{\text{S}})& 0\\
0 & 0 & \rho_\infty
\end{pmatrix},
\end{align}
where the power law resistivity depends on the local current density $\j^\text{S}$ and magnetic flux density $\b^{\text{S}}$ in the HTS layer, that have to be expressed in terms of the macroscale fields $\j$ and $\b$, respectively.

\section{Model Comparison in Axial Field (2D-axi)}\label{sec_resultAxial}

In this section, we consider the 2D-axisymmetric model for the axial configuration. We compare the numerical results and performance of the layered and homogenized models with three different formulations: the \hpfcOnly, \hbfcOnly, and \ajfcOnly-formulations. Technical details about these formulations can be found in~\cite{dular2023standard}.

The \hpf is a standard formulation for problems with HTS and the other two are mixed formulations designed for treating simple HTS-FM hybrid geometries in an efficient manner~\cite{dular3Dmagnet}. We investigate here whether they are relevant choices for the more involved stacked-tape geometry.

Note that other methods such as the homogeneous \tafOnly~\cite{grilli2020electromagnetic} or thin-shell \hpfOnly~\cite{desousa2021thin} are relevant possibilities to consider in further works.

The characteristic mesh size is chosen to scale linearly with a number $\alpha$ that defines the level of refinement. \textit{Coarse}, \textit{medium}, and \textit{fine} levels are defined by $\alpha = 4$, $2$ and $1$, respectively. The coarse meshes associated with $\alpha = 4$ for both models are illustrated in Figs.~\ref{geometry_presentation}(a) and \ref{homogeneousModel}(a).

We model the response from $0$ to $670$~mT at a constant rate of 5~mT/s with steps of $16.75$~mT. In case of non-convergence, the steps are reduced using an adaptive time-stepping procedure~\cite{dular2023standard}.

\subsection{Discretization of the Layered Model}

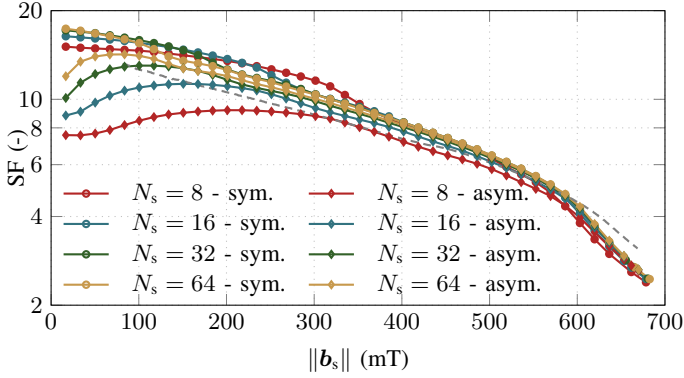
\begin{figure}[h!]
\centering
   \tikzsetnextfilename{SF_axial_influenceOfN1}
		\begin{tikzpicture}[trim axis left, trim axis right][font=\small]
 	\begin{semilogyaxis}[
    width=1.1\linewidth,
    height=5.5cm,
    grid = both,
    grid style = dotted,
    xmin=0, 
    xmax=700,
    ymin=2, 
    ymax=20,
    ytick={2,4,6,8,10,20},
    yticklabels={$2$,$4$,$6$,$8$,$10$,$20$},
	xlabel={$\|\bs\|$ (mT) },
    ylabel={SF (-)},
    ylabel style={yshift=-2.5em},
    xlabel style={yshift=0.2em},
    legend columns=4,
    transpose legend,
    legend style={at={(0.01, 0.0)}, cells={anchor=west}, anchor=south west, draw=none, fill opacity=0, text opacity = 1,style={column sep=0.22cm}, inner sep=1.8pt}
    ]
    \addplot[myred, mark=*, thick, mark options={scale=0.6, style={solid}}] 
    table[x=b,y=SF]{data_stackedTapes/st_axi_deta_h_08_mm1_HTS_bottom.txt};
    \addplot[myblue, mark=*, thick, mark options={scale=0.6, style={solid}}] 
    table[x=b,y=SF]{data_stackedTapes/st_axi_deta_h_16_mm1_HTS_bottom.txt};
        \addplot[mygreen, mark=*, thick, mark options={scale=0.6, style={solid}}] 
    table[x=b,y=SF]{data_stackedTapes/st_axi_deta_h_32_mm1_HTS_bottom.txt};
            \addplot[myorange, mark=*, thick, mark options={scale=0.6, style={solid}}] 
    table[x=b,y=SF]{data_stackedTapes/st_axi_deta_h_64_mm1_HTS_bottom.txt};
        \addplot[myred, mark=diamond*, thick, mark options={scale=0.65, style={solid}}] 
    table[x=b,y=SF]{data_stackedTapes/st_axi_deta_h_08_mm1.txt};
    \addplot[myblue, mark=diamond*, thick, mark options={scale=0.65, style={solid}}] 
    table[x=b,y=SF]{data_stackedTapes/st_axi_deta_h_16_mm1.txt};
        \addplot[mygreen, mark=diamond*, thick, mark options={scale=0.65, style={solid}}] 
    table[x=b,y=SF]{data_stackedTapes/st_axi_deta_h_32_mm1.txt};
            \addplot[myorange, mark=diamond*, thick, mark options={scale=0.65, style={solid}}] 
    table[x=b,y=SF]{data_stackedTapes/st_axi_deta_h_64_mm1.txt};
        \addplot[gray, densely dashed, thick]
    table[x=b,y=SF]{data_stackedTapes/SF_77K_sampleB_axial_highField.txt};
    \legend{$N_{\text{s}}=8$ - sym.,$N_{\text{s}}=16$ - sym.,$N_{\text{s}}=32$ - sym.,$N_{\text{s}}=64$ - sym.,$N_{\text{s}}=8$ - asym.,$N_{\text{s}}=16$ - asym.,$N_{\text{s}}=32$ - asym.,$N_{\text{s}}=64$ - asym.}
    \end{semilogyaxis}
	\end{tikzpicture}%
		\vspace{-0.2cm}
\caption{Influence of $N_{\text{s}}$ in the layered model in the axial configuration. Results from the \hpf and $\alpha = 1$. The dashed gray curve corresponds to the experimental measurements.}
\label{SF_axial_influenceOfN1}
\end{figure}

To assess the influence of $N_{\text{s}}$ on the SF values in the layered model, we perform a series of simulations with an increasing number of tapes, as shown in Fig.~\ref{SF_axial_influenceOfN1}. In addition to the asymmetric geometry of Fig.~\ref{geometry_presentation}, we consider a symmetric variation where the tapes are stacked upside-down for $z<0$, so that we have a plane symmetry with respect to $z$. We first consider a fine mesh ($\alpha = 1$) and the \hpfOnly.

Figure~\ref{SF_axial_influenceOfN1} shows that the results are strongly affected by $N_{\text{s}}$, especially at low field amplitudes and with the asymmetric model. A sufficient number $N_\text{s}$ is necessary to obtain reliable SF predictions. 


Results are also sensitive to the symmetry of the model. In the asymmetric case, magnetic flux lines are channelled into the FM layer at the bottom of the stack, whereas in the symmetric case the HTS layer repels them away from the cylinder. This influences the field at the center of the stack and explains why SF values are lower in the asymmetric case.

The mesh size should be chosen carefully at low fields. Indeed, for low fields, the screening currents only circulate over a thin region at the outer edges of the HTS tapes, as shown in the upper-left part of Fig.~\ref{axial_detailed_homogeneous_jtheta_comparison}. For an accurate evaluation of the SF in that regime, the associated penetrated region should be meshed with enough elements. With $\alpha = 1$, we computed that the relative difference on the SF values between the three formulations is less than 10$\%$ at the lowest considered field ($16.75$ mT) for $N_{\text{s}} =64$.

\subsection{Comparison with the Homogenized Model}

\begin{figure}[h!]
\centering
   \tikzsetnextfilename{SF_axial_modelComparison}
		\begin{tikzpicture}[trim axis left, trim axis right][font=\small]
 	\begin{semilogyaxis}[
    width=1.05\linewidth,
    height=4.8cm,
    grid = both,
    grid style = dotted,
    xmin=0, 
    xmax=700,
    ymin=2, 
    ymax=20,
    ytick={2,4,6,8,10,20},
    yticklabels={$2$,$4$,$6$,$8$,$10$,$20$},
	xlabel={$\|\bs\|$ (mT) },
    ylabel={SF (-)},
    ylabel style={yshift=-2.5em},
    xlabel style={yshift=0.2em},
    legend columns=1,
    transpose legend,
    legend style={at={(0.01, 0.01)}, cells={anchor=west, xshift=-0.3cm}, anchor=south west, draw=none, fill opacity=0, text opacity = 1, style={column sep=0.22cm}, inner sep=1pt}
    ]
        \addplot[mygreen, thick] 
    table[x=b,y=SF]{data_stackedTapes/st_axi_homo_hb_mm0p5_fixed.txt};
     \addplot[myred, thick] 
    table[x=b,y=SF]{data_stackedTapes/st_axi_deta_h_64_mm1_HTS_bottom.txt};
      \addplot[myorange, thick] 
    table[x=b,y=SF]{data_stackedTapes/st_axi_deta_h_64_mm1.txt};
        \addplot[gray, densely dashed, thick]
    table[x=b,y=SF]{data_stackedTapes/SF_77K_sampleB_axial_highField.txt};
    \legend{Homogenized model, Layered model - $N_{\text{s}} = 64$ - symmetric, Layered model - $N_{\text{s}} = 64$ - asymmetric, Experimental measurements}
    \end{semilogyaxis}
	\end{tikzpicture}%
		\vspace{-0.2cm}
\caption{Shielding factors from the \hpf on the layered and homogenized models in the axial configuration. Fine mesh ($\alpha =1$).}
\label{SF_axial_modelComparison}
\end{figure}
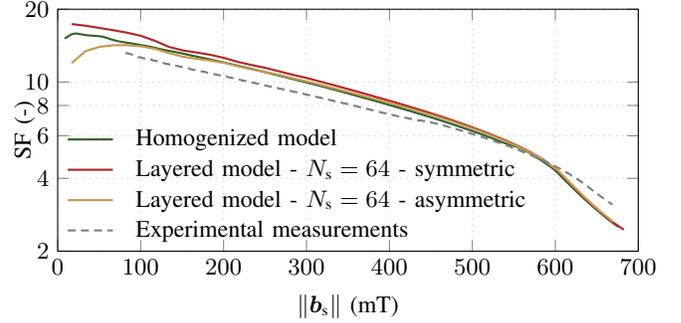

\begin{figure}[h!]
            \centering 
		\includegraphics[width=0.95\linewidth]{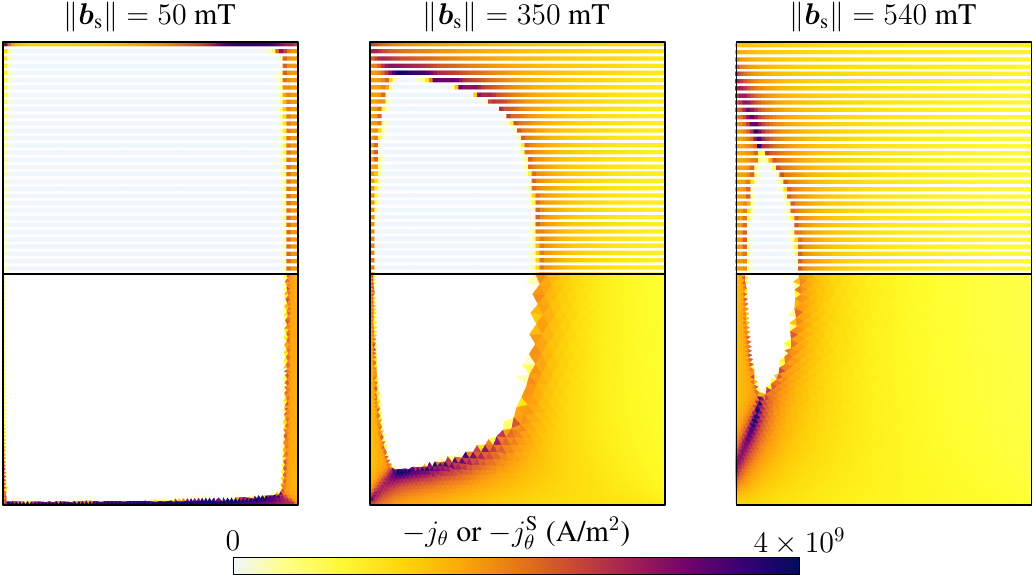}
\caption{Azimuthal current density obtained by the two models in the axial configuration for three values of the applied field. (Up) Layered model with $N_{\text{s}} = 64$ and the \hpfOnly. The HTS layer have been thickened for clarity purposes. (Down) Homogenized model with the \hbfOnly.}
\label{axial_detailed_homogeneous_jtheta_comparison}
\end{figure}

As with the layered model, the mesh size in the homogenized model has to be chosen in accordance with the desired accuracy at low fields. With $\alpha = 1$, the relative difference on SF computed by the different formulations at $16.75$~mT is 21\%, and drops below 10\% for higher values of the applied field. The agreement between the homogenized and layered models is illustrated in Figs.~\ref{SF_axial_modelComparison} and \ref{axial_detailed_homogeneous_jtheta_comparison} for $\alpha =1$. Note that both are also in fair agreement with the experimental measurements.

\subsection{Comparison of the Numerical Efficiencies}

For a given discretization level, both models give results of comparable accuracy with the three formulations. However, they are not equivalent in terms of computational cost. The computational time (CPU time) is mainly determined by two indicators: the number of degrees of freedom (DOFs) and the total number of iterations. Key performance figures are summarized in Table~\ref{table_results_axial}.

\begin{table}[h]
\caption{Performance of the models in the axial case (2D-axi).}
\vspace{-0.35cm}
\begin{center}
\begin{tabular}{l l r c r}
\hline
\multicolumn{1}{c}{Model} & \multicolumn{1}{c}{Formul.} & \# DOFs & \# iterations (\# t.s.) & \multicolumn{1}{c}{Total time} \\
\hline
\multirow{3}{*}{$N_{\text{s}} = 16$} & \hpfc & 44\,516 & 1\,538 (57) & 51 m\ \ ~\\
 & \hbfc & 78\,298 & 1\,100 (42) & 49 m\ \ ~\\
 & \ajfc & 45\,622 & 1\,233 (40) & 38 m\ \ ~\\
 \hline
\multirow{3}{*}{$N_{\text{s}} = 32$} & \hpfc & 51\,261 & 1\,599 (64) & 1 h 05 m\ \ ~\\
 & \hbfc & 85\,237 & 1\,431 (57) & 1 h 12 m\ \ ~\\
 & \ajfc & 53\,679 & 1\,361 (43) & 51 m\ \ ~\\
  \hline
  \multirow{4}{*}{$N_{\text{s}} = 64$} & \hpfc & 55\,925 & \ \ 2\,536 (112) & 1 h 51 m\ \ ~\\
 & \hbfc & 92\,290 &  \ \ 2\,491 (106) & 2 h 22 m\ \ ~\\
 & \ajfc \tiny{($\alpha = 1$)} & 60\,967 & \footnotesize{Not converged} & \multicolumn{1}{c}{/}\\
 & \textcolor{darkgray}{\ajfc \tiny{($\alpha = 0.5$)}}& \textcolor{darkgray}{192\,916} & \textcolor{darkgray}{2\,419 (54)} & \textcolor{darkgray}{6 h 59 m}\ \ ~\\
  \hline
  \hline
\multirow{3}{*}{Homog.} & \hpfc & 49\,068 & 2\,522 (95) & 1 h 42 m\ \ ~\\
 & \hbfc & 62\,807 & \ \ 1\,828 (124) & 2 h 09 m\ \ ~\\
 & \ajfc & 44\,498 & 1\,176 (40) & 48 m\ \ ~\\
   \hline
\end{tabular}
\end{center}
Performance of the different approaches and formulations, on a fine mesh resolution with $\alpha = 1$, except for the \ajf with $N_{\text{s}} = 64$, for which $\alpha = 0.5$. The layered model is symmetric and the homogenized model accounts for the $z$-plane symmetry. Simulation up to $\|\bs\| = 670$ mT in the axial configuration, with a minimum of 40 time steps. The actual number of time steps (t.s.) resulting from the adaptive time-stepping procedure is given within parenthesis in the table. The CPU times are for a single AMD EPYC Rome CPU at 2.9 GHz.
\label{table_results_axial}
\end{table}

Among the different formulations, the \ajf demonstrates the highest overall efficiency in terms of CPU time, but it suffers from solving difficulties with the default linear solver in MUMPS \cite{amestoy2000mumps} for high values of $N_\text{s}$ if the mesh is not sufficiently fine. These difficulties already appear for $N_\text{s} = 16$ for $\alpha = 2$, and were also observed in earlier works in other situations involving global variables~\cite{dular2023standard}.

None of the two models, layered or homogenized, significantly outperforms the other in the axial configuration. The computational cost associated with the implicit equation resolution in the homogenized model only accounts for approximately 6\% of the total cost of the simulation, and does not disqualify the approach. Both approaches are equally valid. Most importantly, the mesh discretization should be fine enough for the desired accuracy at low fields.

\section{Model Comparison in Transverse Field (3D)}\label{sec_resultTransverse}

For the transverse configuration, we consider a mesh whose characteristic size scales linearly a multiplier $\alpha$ as in the 2D model. The coarse meshes associated with $\alpha = 4$ for both models are represented in Figs.~\ref{geometry_presentation}(b) and \ref{homogeneousModel}(b) (the outside of the stack is not represented for clarity).

Note that aligning the elements with the principal directions of anisotropy and using prismatic elements was found to significantly help the resolution for the homogenized model.

We model the magnetic response of the system from $0$ to $60$~mT, with steps of $1.5$~mT, and use an adaptive time-stepping procedure in the case of non-convergence.

\subsection{Discretization of the Layered Model}

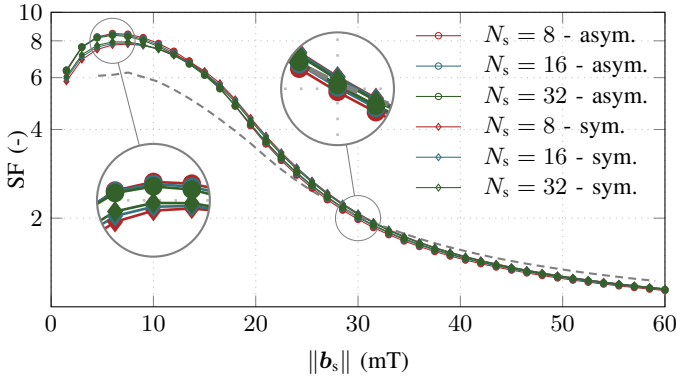
\begin{figure}[h!]
\centering
   \tikzsetnextfilename{SF_transverse_influenceOfN1}
		\begin{tikzpicture}[trim axis left, trim axis right,spy using outlines={circle, magnification=2.5, connect spies}][font=\small]
 	\begin{semilogyaxis}[
    width=1.1\linewidth,
    height=5.5cm,
    grid = both,
    grid style = dotted,
    xmin=0, 
    xmax=60,
    ymin=1, 
    ymax=10,
    ytick={2,4,6,8,10,20},
    yticklabels={$2$,$4$,$6$,$8$,$10$,$20$},
	xlabel={$\|\bs\|$ (mT) },
    ylabel={SF (-)},
    ylabel style={yshift=-2.5em},
    xlabel style={yshift=0.2em},
    legend columns=1,
    legend style={at={(0.99, 0.99)}, cells={anchor=west}, anchor=north east, draw=none, fill opacity=0, text opacity = 1,style={column sep=0.22cm}, inner sep=1.8pt}
    ]
    \addplot[myred, mark=*, mark options={scale=0.6, style={solid}}] 
    table[x=b,y=SF]{data_stackedTapes/st_tr_08_m1_asym.txt};
    \addplot[myblue, mark=*, mark options={scale=0.6, style={solid}}] 
    table[x=b,y=SF]{data_stackedTapes/st_tr_16_m1_asym.txt};
        \addplot[mygreen, mark=*, mark options={scale=0.6, style={solid}}] 
    table[x=b,y=SF]{data_stackedTapes/st_tr_32_m1_asym.txt};
        \addplot[myred, mark=diamond*, mark options={scale=0.65, style={solid}}] 
    table[x=b,y=SF]{data_stackedTapes/st_tr_08_m1_sym.txt};
    \addplot[myblue, mark=diamond*, mark options={scale=0.65, style={solid}}] 
    table[x=b,y=SF]{data_stackedTapes/st_tr_16_m1_sym.txt};
        \addplot[mygreen, mark=diamond*, mark options={scale=0.65, style={solid}}] 
    table[x=b,y=SF]{data_stackedTapes/st_tr_32_m1_sym.txt};
        \addplot[gray, densely dashed, thick]
    table[x=b,y=SF]{data_stackedTapes/SF_77K_sampleB_transverse.txt};
    \legend{$N_{\text{s}}=8$ - asym.,$N_{\text{s}}=16$ - asym.,$N_{\text{s}}=32$ - asym.,$N_{\text{s}}=8$ - sym.,$N_{\text{s}}=16$ - sym.,$N_{\text{s}}=32$ - sym.}
     \coordinate (spypoint) at (axis cs: 6,8);
	\coordinate (magnifyglass) at (axis cs: 10,2.3);
	     \coordinate (spypoint_2) at (axis cs: 30,2);
	\coordinate (magnifyglass_2) at (axis cs: 28,5.5);
    \end{semilogyaxis}
    \spy [gray, size=1.5cm] on (spypoint)
	in node[fill=white] at (magnifyglass);
	    \spy [gray, size=1.5cm] on (spypoint_2)
	in node[fill=white] at (magnifyglass_2);
	\end{tikzpicture}%
		\vspace{-0.2cm}
\caption{Influence of $N_{\text{s}}$ in the layered model in the transverse configuration, for a fine mesh ($\alpha = 1$) and the \hpfOnly. Asymmetric and symmetric geometries are considered. The dashed gray curve corresponds to the experimental measurements.}
\label{SF_transverse_influenceOfN1}
\end{figure}

As shown in Fig.~\ref{SF_transverse_influenceOfN1}, in the transverse case, the SF values are much less sensitive to $N_\text{s}$ than in the axial case. Similarly, the symmetry of the stack hardly affects the solution. In the following, we choose the symmetric model and $N_\text{s} =16$.

\begin{figure}[h!]
\centering
   \tikzsetnextfilename{SF_transverse_simple_meshConvergence}
		\begin{tikzpicture}[trim axis left, trim axis right,spy using outlines={circle, magnification=2.5, connect spies}][font=\small]
 	\begin{semilogyaxis}[
    width=1.1\linewidth,
    height=5.5cm,
    grid = both,
    grid style = dotted,
    xmin=0, 
    xmax=60,
    ymin=1, 
    ymax=10,
    ytick={2,4,6,8,10,20},
    yticklabels={$2$,$4$,$6$,$8$,$10$,$20$},
	xlabel={$\|\bs\|$ (mT) },
    ylabel={SF (-)},
    ylabel style={yshift=-2.5em},
    xlabel style={yshift=0.2em},
    legend style={at={(0.99, 0.99)}, cells={anchor=west}, anchor=north east, draw=none, fill opacity=0, text opacity = 1,style={column sep=0.22cm}, inner sep=1.8pt}
    ]
            \addplot[hbf, mark=diamond*, mark options={scale=0.65, style={solid}}] 
    table[x=b,y=SF]{data_stackedTapes/st_tr_16_hb_eighth_m4.txt};
    \addplot[hbf, mark=square*, thick, mark options={scale=0.6, style={solid}}] 
    table[x=b,y=SF]{data_stackedTapes/st_tr_16_hb_eighth_m2.txt};
        \addplot[hbf, mark=*, thick, mark options={scale=0.6, style={solid}}] 
    table[x=b,y=SF]{data_stackedTapes/st_tr_16_hb_eighth_m1.txt};
        \addplot[ajf, mark=diamond*, mark options={scale=0.65, style={solid}}] 
    table[x=b,y=SF]{data_stackedTapes/st_tr_16_aj_eighth_m4.txt};
    \addplot[ajf, mark=square*, thick, mark options={scale=0.6, style={solid}}] 
    table[x=b,y=SF]{data_stackedTapes/st_tr_16_aj_eighth_m2.txt};
        \addplot[ajf, mark=*, thick, mark options={scale=0.6, style={solid}}] 
    table[x=b,y=SF]{data_stackedTapes/st_tr_16_aj_eighth_m1_adapted.txt};
        \addplot[gray, densely dashed, thick]
    table[x=b,y=SF]{data_stackedTapes/SF_77K_sampleB_transverse.txt};
    \legend{$h$-$\phi$(-$b$) - coarse,$h$-$\phi$(-$b$) - medium, $h$-$\phi$(-$b$) - fine, \ajfc - coarse, \ajfc - medium, \ajfc - fine}
    \end{semilogyaxis}
	\end{tikzpicture}%
		\vspace{-0.2cm}
\caption{Shielding factors computed by three formulations on the layered model ($N_{\text{s}}=16$) in the transverse configuration, for three discretization levels ($\alpha = 4$, $2$, and $1$). Curves for the \hpfc and \hbfcOnly-formulations are visually indistinguishable. The case $\alpha = 1$ for the \ajf was stopped at 7.5 mT after 30 hours of CPU time. The dashed gray curve corresponds to the experimental measurements.}
\label{SF_transverse_simple_meshConvergence}
\end{figure}
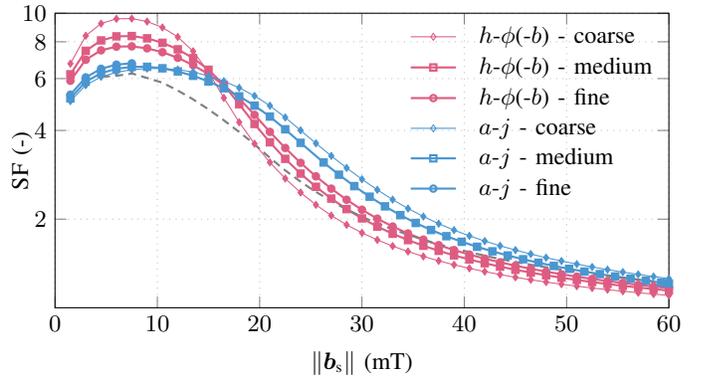

The SF values obtained with the three formulations for different discretization levels are given in Fig.~\ref{SF_transverse_simple_meshConvergence}. The solutions approach each other with mesh refinement, but non-negligible changes are still observed between the medium and fine meshes. As shown in Table~\ref{table_results_transverse}, the associated simulations already exceed $7$ hours of CPU time, which illustrates how challenging this problem is in 3D.

\subsection{Comparison with the Homogenized Model}

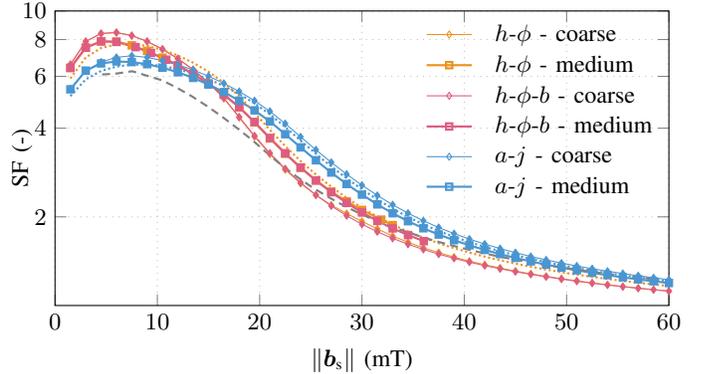
\begin{figure}[h!]
\centering
   \tikzsetnextfilename{SF_transverse_homogeneous_meshConvergence}
	\begin{tikzpicture}[trim axis left, trim axis right,spy using outlines={circle, magnification=2.5, connect spies}][font=\small]
 	\begin{semilogyaxis}[
    width=1.1\linewidth,
    height=5.5cm,
    grid = both,
    grid style = dotted,
    xmin=0, 
    xmax=60,
    ymin=1, 
    ymax=10,
    ytick={2,4,6,8,10,20},
    yticklabels={$2$,$4$,$6$,$8$,$10$,$20$},
	xlabel={$\|\bs\|$ (mT) },
    ylabel={SF (-)},
    ylabel style={yshift=-2.5em},
    xlabel style={yshift=0.2em},
    legend style={at={(0.99, 0.99)}, cells={anchor=west}, anchor=north east, draw=none, fill opacity=0, text opacity = 1,style={column sep=0.22cm}, inner sep=1.8pt}
    ]
        \addplot[hf, mark=diamond*, mark options={scale=0.65, style={solid}}] 
    table[x=b,y=SF]{data_stackedTapes/st_tr_ho_h_m4_prisms.txt};
    \addplot[hf, mark=square*, thick, mark options={scale=0.6, style={solid}}] 
    table[x=b,y=SF]{data_stackedTapes/st_tr_ho_h_m2_prisms.txt};
       \addplot[hbf, mark=diamond*, mark options={scale=0.65, style={solid}}] 
    table[x=b,y=SF]{data_stackedTapes/st_tr_ho_hb_m4_prisms.txt};
    \addplot[hbf, mark=square*, thick, mark options={scale=0.6, style={solid}}] 
    table[x=b,y=SF]{data_stackedTapes/st_tr_ho_hb_m2_prisms_new_2.txt};
        \addplot[ajf, mark=diamond*, mark options={scale=0.65, style={solid}}] 
    table[x=b,y=SF]{data_stackedTapes/st_tr_ho_aj_m4_prisms.txt};
    \addplot[ajf, mark=square*, thick, mark options={scale=0.6, style={solid}}] 
    table[x=b,y=SF]{data_stackedTapes/st_tr_ho_aj_m2_prisms.txt};
                \addplot[hf, densely dotted, thick] 
    table[x=b,y=SF]{data_stackedTapes/st_tr_16_h_eighth_m1.txt};
        \addplot[ajf, densely dotted, thick] 
    table[x=b,y=SF]{data_stackedTapes/st_tr_16_aj_eighth_m2.txt};
        \addplot[gray, densely dashed, thick]
    table[x=b,y=SF]{data_stackedTapes/SF_77K_sampleB_transverse.txt};
   \legend{\hpfc - coarse,\hpfc - medium,  \hbfc - coarse, \hbfc - medium, \ajfc - coarse, \ajfc - medium}
    \end{semilogyaxis}
	\end{tikzpicture}%
		\vspace{-0.2cm}
\caption{Shielding factors from the \hpfcOnly, \hbfcOnly, and \ajfcOnly-formulations on the homogenized model in the transverse configuration, for different discretization levels. The dotted curves are given for comparison with the layered model ($h$-$\phi$ fine and $a$-$j$ medium). The dashed gray curve corresponds to the experimental measurements.}
\label{SF_transverse_homogeneous_meshConvergence}
\end{figure}

\begin{figure}[h!]
            \centering 
		\includegraphics[width=\linewidth]{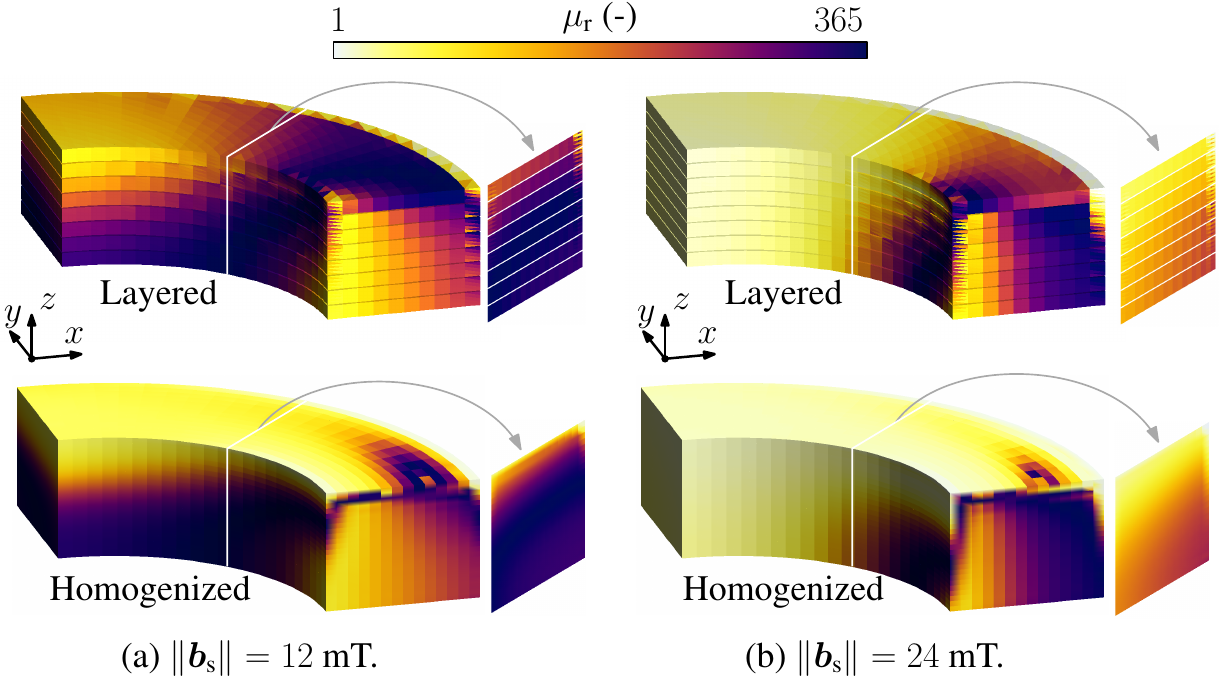}
\caption{Relative permeability obtained by the two models in the transverse configuration and the \hpfOnly, with the medium mesh resolution ($\alpha = 2$) for two values of applied field. The solution in the plane $x=y$ is indicated by the gray arrows. (Up) Layered model, $N_\text{s} = 16$. (Down) Homogenized model, the permeability is the scalar value $\mur(\h^\text{F})$.}
\label{stack_mur_comparison_transverse}
\end{figure}

The homogenized model produces solutions of similar quality compared to the layered model, as shown in Figs.~\ref{SF_transverse_homogeneous_meshConvergence} and \ref{stack_mur_comparison_transverse}. Solutions may differ locally, e.g., on the top of the stack of tapes for the relative permeability, but the overall field distributions and SF values are in good agreement.

\subsection{Comparison of the Numerical Efficiencies}

In contrast to the case of the axial configuration, significant differences in numerical efficiency are observed in the transverse case. Performance figures are gathered in Table~\ref{table_results_transverse}.

\begin{table}[h]
\caption{Performance of the models in the transverse case (3D).}
\vspace{-0.35cm}
\begin{center}
\begin{tabular}{c l r c r r r}
\hline
\multicolumn{1}{c}{Model} & \multicolumn{1}{c}{Formul.} & \# DOFs & \# it. (\# t.s.) & \multicolumn{1}{c}{Total time} \\
\hline
\multirow{3}{*}{\shortstack{Layered\\ Coarse}} & \hpfc & 3\,579 & 413 (40) & 3 m 46\ \ ~\\
 & \hbfc & 10\,419 & 165 (40) & 1 m 37\ \ ~\\
 & \ajfc & 18\,224 & 172 (40) & 3 m 50\ \ ~\\
  \hline
\multirow{3}{*}{\shortstack{Layered\\ Medium}} & \hpfc & 23\,511 & 471 (40) & 35 m 50\ \ ~\\
 & \hbfc & 84\,015 & 222 (40) & 34 m 40\ \ ~\\
 & \ajfc & 120\,716 & 765 (43) & 3 h 49 m\ \quad\ \ ~\\
   \hline
\multirow{3}{*}{\shortstack{Layered\\ Fine}}  & \hpfc & 151\,676 & 524 (40) & 7 h 23 m\ \quad\ \ ~\\
 & \hbfc & 631\,204 & 278 (40) & 13 h 46 m\ \quad\ \ ~\\
 & \ajfc & 823\,073 & \textcolor{darkgray}{120 (6)}\ ~ & \multicolumn{1}{c}{\textcolor{darkgray}{$>$30 h\, \,($7.5$ mT)}}\\
   \hline
  \hline
  \multirow{3}{*}{\shortstack{Homog.\\ Coarse}}  & \hpfc & 3\,406 & 560 (40) & 4 m 57\ \ ~\\
 & \hbfc & 6\,118 & 295 (40) & 4 m 17\ \ ~\\
 & \ajfc & 11\,960 & 196 (40) & 3 m 36\ \ ~\\
  \hline
\multirow{3}{*}{\shortstack{Homog.\\ Medium}}& \hpfc & 28\,636 & \textcolor{darkgray}{3\,383 (38)}\ ~  &  \multicolumn{1}{c}{\textcolor{darkgray}{$>$\, 6 h\, \,($33$ mT)}}~\\
 & \hbfc & 53\,260 & \textcolor{darkgray}{4\,822 (252)} & \multicolumn{1}{c}{\textcolor{darkgray}{$>$48 h\, \,($36$ mT)}}~\\
 & \ajfc & 84\,938 & 307 (40) & 3 h 31 m\ \quad\ \ ~\\
\hline
\end{tabular}
\end{center}
Performance of the different approaches and formulations. Simulation up to $\|\bs\| = 60$ mT in the transverse configuration, with a minimum of 40 time steps. The layered model is symmetric with $N_\text{s} = 16$ and the homogenized model accounts for the $z$-plane symmetry. The actual number of time steps (t.s.) resulting from the adaptive time-stepping procedure is given within parenthesis in the fourth column of the table. Gray values are associated with simulations that did not end properly: either because of exceedingly slow convergence (\ajfc and \hbfcOnly), or because of iteration cycles (\hpfcOnly); the field values within parenthesis in the last column indicates where the simulation was stopped. The CPU times are for a AMD EPYC Rome CPU at 2.9 GHz.
\label{table_results_transverse}
\end{table}

With the layered model, the \hbf converges in a smaller number of iterations than the \hpfOnly. This is directly related to the fact that the reluctivity can be more efficiently handled than the permeability by iterative methods~\cite{dular2019finite}. However, the number of DOFs involved with the \hbf is much higher, so that the CPU time associated with the \hbf becomes higher for the fine mesh resolution.

For the layered model, the performance of the \ajf deteriorates with mesh refinement, and does not compete with the other two for similar solution quality. The number of DOFs is higher than for the other formulations in this 3D problem, and the number of iterations significantly increases with finer meshes.

For the homogenized model, mixed formulations perform better than the \hpf at the coarse discretization level, thanks to a significant gain in the number of iterations. 

At the medium discretization level, the \ajf is the only formulation that converges with only 40 time steps. The \hpf enters iteration cycles at an applied field of $33$ mT, even with a hybrid Picard-Newton-Raphson iterative technique~\cite{dular2023standard}. The adaptive time-stepping procedure does not help to avoid these cycles in this case and convergence fails.

For the \hbf at medium level, the simulation does not end within 48 hours of CPU time, whereas for the layered model at a comparable discretization level, the complete simulation is achieved in less than an hour. Contrary to the situation with the \hpfOnly, no iteration cycle is observed, but very small time steps are necessary to avoid divergence of the Newton-Raphson iterations. We can explain these solving difficulties by the combination of two factors: anisotropic materials properties, and strongly nonlinear constitutive laws in a hybrid magnetic-conducting region. This seems to disqualify the \hbf in 3D for this geometry.

For the fine discretization level on the homogenized model (not represented in the table), the \hpfc and \hbfcOnly-formulations face the same difficulties as in the medium level, and the \ajf becomes exceedingly slow, as is the case in the layered model at the fine level.

The performance figures in Table \ref{table_results_transverse} lead to the conclusion that the layered model should be preferred to the homogenized model in the 3D case. The computational cost of the latter model rapidly increases with mesh refinement and convergence difficulties are encountered with the \hpfc and \hbfcOnly-formulations, whereas in the layered model, these two formulations lead to robust resolutions. At fine discretization levels, we recommend the \hpf together with the layered model.

\section{Conclusion}\label{sec_conclusion}

We compared the numerical performance of different modelling approaches for computing the shielding effectiveness of a stack of HTS tape annuli. Conclusions for the axial (2D-axi) and the transverse (3D) configurations are not identical. In the axial case, the layered and homogenized models demonstrate comparable performance. Among the tested formulations, the \ajf is the most efficient, although one may expect solving difficulties with it, depending on the linear solver.

In the transverse case, the homogenized model is noticeably slower than the layered model: the \hpf faces iteration cycles that are not easy to avoid, the \hbf requires very small time steps to converge, and the \ajf is extremely expensive with mesh refinement. On the contrary, the layered model allows for robust and efficient resolutions, with the \hbf for coarse discretization levels, and with the \hpf for finer ones. The \ajf does not compete with these formulations in 3D.

\section*{Acknowledgment}

Computational resources have been provided by the Consortium des Équipements de Calcul Intensif (CÉCI), funded by the Fonds de la Recherche Scientifique de Belgique (F.R.S.-FNRS) under Grant No. 2.5020.11.

\clearpage
\bibliographystyle{ieeetr}
\bibliography{paperReferences}

\end{document}